# Comparison of different ML methods applied to the classification of events with $t\bar{t}$ in the final state at the ATLAS experiment

**[1]Samuel Campo Martínez, [1]José  Salt, [1]Santiago González de la Hoz,[2]Miguel Villaplana**

[1] Instituto de Física Corpuscular (IFIC), University of Valencia and CSIC, Valencia, Spain

[2] Istituto Nazionale di Fisica Nucleare (INFN), Sezione di Milano, Italy

ABSTRACT

This contribution describes the experience with the application of different Machine Learning (ML) techniques to a physics analysis case. The use case chosen is the classification of top-antitop events coming from BSM or from SM using data from a repository of simulated events. The features of these events are represented by their kinematic observables.

The initial objective was to compare different ML methods in order to see whether they can lead to an improvement in the classification, but the work has also helped us to test many variations in the methods by changing hyper-parameters, using different optimisers, ensembles, etc. With this information we have been able to conduct a comparative study that is useful for ensuring as complete control as possible of the methodology.

PRESENTED AT

Connecting the Dots and Workshop on Intelligent Trackers (CTD/WIT 2019)
Instituto de Física Corpuscular (IFIC), Valencia, Spain
April 2-5, 2019

## 1. INTRODUCTION

Machine Learning (ML) classifiers have been applied to many physics analysis in LHC experiments (see ref [1]). Our natural physical case for applying them is the search for a new particle X of unknown mass which is a resonance that decays to a top-antitop ($t\bar{t}$) pair produced in collisions of proton-proton at 8 TeV in ATLAS. The main goal is to train a classifier enhancing the selection efficiency of events corresponding to new physics and the rejection of events corresponding to Standard Model (SM) processes with $t\bar{t}$ in the final state (see Fig. 1).

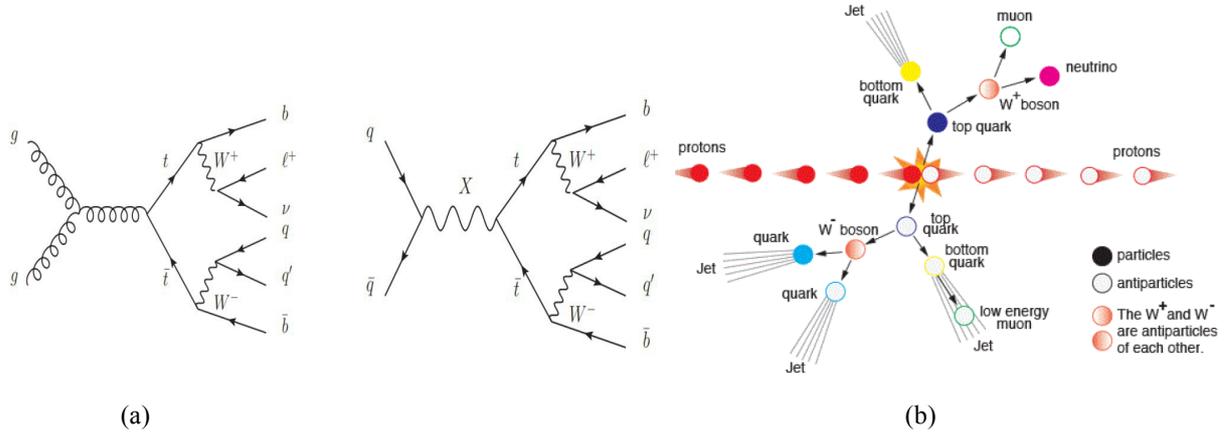

(a)                                               (b)

Figure 1. (a) Feynman Diagrams: Standard Model (background) and Beyond the Standard Model (signal). (b) Decays schema of final state.

In Figure 2-b the anatomy of top decays for a given event is shown. Taking into account the particles in the final state, one can consider the kinematical variables: transverse momentum, pseudorapidity and azimuth of the lepton (muon or electron); transverse momenta, pseudorapidity and azimuth of the four more energetic leading jets; b tagging of these four jets and the transverse momentum and azimuth of missing transverse energy. Moreover, five more variables corresponding to the invariant mass of W decay, of the top and antitop decays and the overall $t\bar{t}$ system.

## 2 INPUT SIMULATED DATA, DISCRIMINATING VARIABLES AND VARIABLE CORRELATIONS

The studies have been done using simulated data generated with MADGRAPH and Pythia from a dataset repository managed by a ATLAS member group [2]. 10 million events have been analysed. Signal datasets have been produced with 5 values of the resonance mass (500 GeV, 1250 GeV, 750 GeV, 1000 GeV, 1500 GeV) and there is another dataset with background events (SM events). For each event, there are 21 low-level features (the variables described before) plus 5 high-level features (the invariant mass ones). We can see the distribution of the transverse momentum of the leading jet for different masses of the signal and for the background in Figure 2. An important aspect is to study the correlation between the variables. In Figure 2-b one can see correlation plot for a subset of 6 variables.

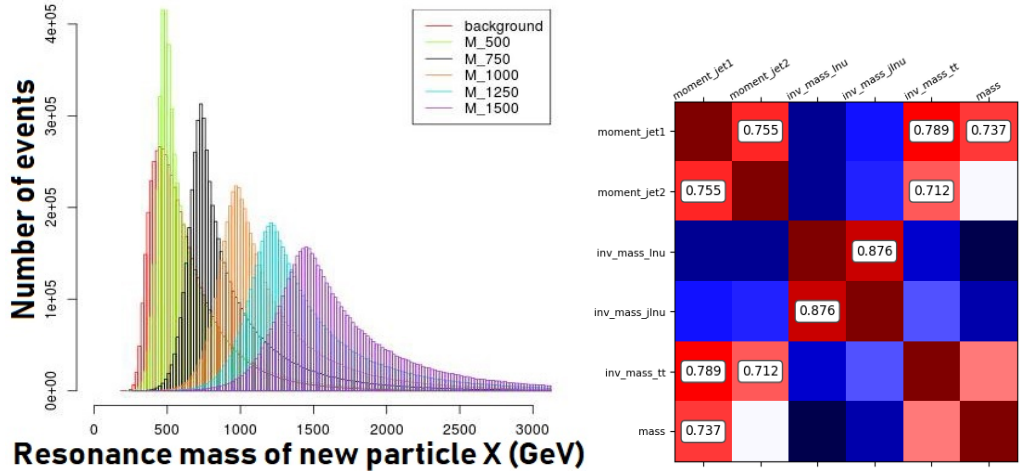

Figure 2: (a) Distribution of $t\bar{t}$ invariant mass for background and signal events generated with different masses, (b) Correlation map for a subset of the variables.

## 3 APPLICATION OF DIFFERENT MACHINE LEARNING METHODS

Several Machine Learning methods are applied to this set of features. We report performance results of three methods: Decision Trees, Logistic Regression and Neural Networks. Each method has been studied doing many variations: changing hyper-parameters, using optimisers, ensembles and lots.

Their classification quality has been evaluated using ROC (Receiving Operating Curves) of True Positive Rate vs False Positive Rate and calculating the AUC (Area Under the Curve) as a parameter of the classification quality. The first method checked is Decision Trees. One of the important aspects is the behaviour of the performance quality (AUC). Samples of different noise ratio (signal/background) are studied to know the effect of noise in the different ML methods.

### 3.1 Extra Trees Classifiers and Logistic Regression.

Extra Trees classifiers are studied and the AUC obtained for different noise percentages is displayed in Figure 3-a. A Logistic Regression method is studied and the AUC obtained for different noise percentages is displayed in Figure 3-b.

### 3.2 Neural Networks.

In our case we have used a neural network with three hidden layers fully interconnected. Each connection has a weight. In the first layer we add the input variables, the hidden layers where import the weights and finally the layer for the output variables. In our architecture the output is composed of two neurones that give us the probability that an event is SM or BSM.

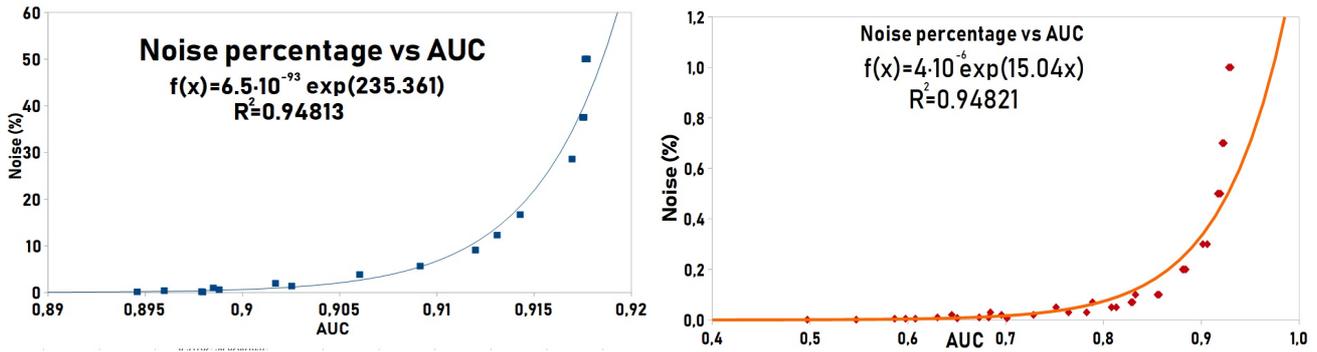

Figure 3: Classification quality (AUC) as a function of the percentage of background in the sample for (a) Logistic Regression (b) Decision Trees.

## 4. CONCLUSIONS

A summary of the results obtained with different ML methods is shown in Table 1.

| Method | Noise (AUC=0.87) | Maximum AUX | Noise (maximum AUC) |
|---|---|---|---|
| Extra Trees | 0.4% | 0.945 | 36.0% |
| Logistic Regression | 0.02% | 0.92 | 29.0% |
| Neural Network | 0.2% | 0.893 | 21.0% |

Table 1: Comparative table for the different ML techniques.

The best AUC with a non-negligible level of noise population is achieved using Extra (Decision) Trees. If we use Logistic Regression, the AUC is also good but with much less noise population. Neural Networks need to have almost half of the noise population to achieve even worse results than Logistic Regression. The results show an unstable behaviour that we are trying to understand.

### ACKNOWLEDGEMENTS

We acknowledge the support of project FPA2016-75141-C1, Ministerio de Economía, Innovación y Competitividad , Spain